\documentclass[sigconf]{acmart}
\usepackage{listings}
\usepackage[htt]{hyphenat}
\usepackage{changepage} 
\usepackage{courier}
\usepackage[T1]{fontenc}
\newcommand{\ie}[1]{\textit{i.e.}}
\newcommand{\eg}[1]{\textit{e.g.}}
\newcommand{\etc}[1]{\textit{etc}}
\newcommand{\etal}[1]{\textit{et al.}}
\newcommand{\platx}[1]{\textit{bauplan}}

\AtBeginDocument{%
  }

\setcopyright{acmlicensed}
\copyrightyear{2018}
\acmYear{2018}
\acmDOI{XXXXXXX.XXXXXXX}

\acmConference[Conference acronym 'XX]{Make sure to enter the correct  conference title from your rights confirmation emai}{June 03--05,
  2018}{Woodstock, NY}
\acmISBN{978-1-4503-XXXX-X/18/06}

\begin{document}

\title{\textit{Bauplan}: zero-copy, scale-up FaaS for data pipelines (pre-print)}

\author{Jacopo Tagliabue}
\affiliation{
 \country{}
  \institution{Bauplan Labs}
}

\author{Tyler Caraza-Harter}
\affiliation{
 \country{}
  \institution{University of Wisconsin-Madison}
}

\author{Ciro Greco}
\affiliation{
 \country{}
  \institution{Bauplan Labs}
}

\renewcommand{\shortauthors}{Tagliabue et al.}

\begin{abstract}
   Chaining functions for longer workloads is a key use case for FaaS platforms in data applications. However, modern data pipelines differ significantly from typical serverless use cases (\eg{}, webhooks and microservices); this makes it difficult to retrofit existing pipeline frameworks due to structural constraints. In this paper, we describe these limitations in detail and introduce \textit{bauplan}, a novel FaaS programming model and serverless runtime designed for data practitioners. \textit{bauplan} enables users to declaratively define functional Directed Acyclic Graphs (DAGs) along with their runtime environments, which are then efficiently executed on cloud-based workers. We show that \textit{bauplan} achieves both better performance and a superior developer experience for data workloads by making the trade-off of reducing generality in favor of data-awareness. 
\end{abstract}




\maketitle

\section{Introduction}
\label{sec:intro}

\begin{figure}
  \centering
  \includegraphics[width=\linewidth]{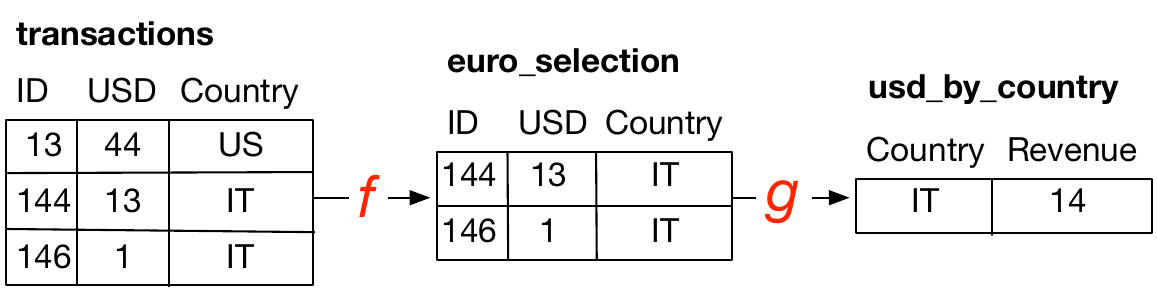}
  \caption{\textit{A DAG of dataframes produced by transformations}: the source dataframe \textit{transactions} is first filtered for European countries (generating \textit{euro\_selection}, then the aggregation of revenues by country is computed as \textit{usd\_by\_country}.}
  \label{fig:dag}
\end{figure}

The growth of Artificial Intelligence and Analytics applications has created a large, fast-growing market for data pipeline tools (USD 10BN/ year, with 22\% CAGR \cite{cargpipeline}). Data pipelines are typically represented as Directed Acyclic Graphs (DAGs), where the nodes are functions that transform, aggregate, or clean the "raw" data for downstream use (Fig.~\ref{fig:dag}).
These functions typically map one or more input dataframes to one or more output dataframes \cite{10.14778/3407790.3407807}. 

This approach allows practitioners to break down complex business logic into modular, reusable components. Data pipeline workloads can be highly variable, even for a single user working on a single DAG: it is common to start with a preliminary run on, say, January data, then scale up to a year, with a corresponding, instantaneous change of dataframe size. In this light, data workloads seem to be a natural fit for Function-as-a-Service (FaaS) platforms designed to efficiently handle bursty, functional, and event-driven tasks. Unfortunately, existing FaaS runtimes fall short in practice as they were primarily designed to support the execution of many simple, independent functions that produce small outputs. Although popular FaaS platforms (\eg{}, AWS~Lambda~\cite{lambda}, Azure Functions~\cite{azure}, and OpenWhisk~\cite{openserwhisk}) have added support for function chaining, their capabilities fall short for data pipelines. It is therefore not surprising that widely used data engineering frameworks (\eg{}, Airflow~\cite{airflow}, Prefect~\cite{prefect}, and Luigi~\cite{luigi}) lack native integration with serverless runtimes.

We group our contributions in two main categories. \textit{First}, based on industry experience, relevant literature, and system traces, we detail the specific demands data pipelines place on FaaS platforms and how current implementations fall short (\S\ref{sec:whynew}). We identify three key areas for improvement: scaling to handle large workloads, efficiently passing intermediate dataframes between functions, and rapidly adapting to developer changes. \textit{Second}, we introduce \platx{}, a specialized FaaS service built for executing data pipelines. \platx{} fills the gap between traditional FaaS platforms and pipeline abstractions, offering a truly serverless experience designed with the needs of data workloads in mind. It features a novel programming model with simple annotations for standard Python or SQL functions, and a runtime enabling platform-level optimizations similar to a database query planner. For example, unlike Lambda's Docker-based execution for custom dependencies, \platx{} provides a declarative API for managing Python packages at the function level, leaving caching and optimization to the platform. In common data science scenarios (Table~\ref{tab:napkin}), \platx{}'s build process is 15x faster than \textit{AWS Lambda} and 7x faster than \textit{Snowpark}.

Unlike general-purpose FaaS platforms that allow any kind of input and output (provided that users manually take care of serialization and triggering), \platx{} stores all intermediate data as (or automatically converted to) \textit{Arrow} tables. Arrow is an open-source in-memory tabular format with a rich ecosystem that enables zero-copy data sharing between nodes on the same host and avoids serialization costs when data is sent across the wire (\S\ref{sec:passing}). Finally, \platx{} planning and scheduling is managed by the provider, but invocations occur on the customer's virtual machines; a single invocation can use nearly all a machine's resources, so \platx{} provides maximal scale up.

\section{Do we need a new FaaS? An industry perspective}
\label{sec:whynew}


The life cycle of data projects is only superficially comparable to traditional software development \cite{https://doi.org/10.48550/arxiv.2209.09125}, with its own pitfalls and failure modes~\cite{datasurvery,Tagliabue2023ReasonableSM}.
In contrast to traditional serverless use cases (\eg{}, endpoints for web applications \cite{sciabarra}), pipelines need support in scaling up, handling intermediate artifacts, and boosting interactivity:

\begin{enumerate}
    \item \textit{scaling up}: horizontal scaling through replicas is not as important as the ability to re-run the same function while (massively) scaling up hardware requirements between executions: recent industry traces place the \textit{p99.9} memory footprint of a function between 50 and 200 GB for most real-world use cases~\cite{Renen2024};
    \item \textit{large intermediate I/O}: functions often deal with large (\textgreater 10 GBs) dataframes  as their inputs and outputs, so the cost of serializing and moving the payload may be significant; 
    \item \textit{fast feedback loop}: unlike most software development projects, data science projects are open-ended and exploratory in nature; if no strategy is known from the start, the key to success is rapidly iterating over a set of hypotheses~\cite{Xin2018HowDI}.
\end{enumerate}

Below, we provide preliminary benchmarks on the cost of containerization (\S\ref{sec:container}) and data movement (\S\ref{sec:passing}) in traditional platforms vs. \platx{}. Here, we offer a brief comparison of commercial and open source FaaS limitations \textit{vis-à-vis} the peculiarities of data pipelines. On the commercial side, we picked the two largest by market share, \textit{AWS Lambda} and \textit{Azure Functions} \cite{azure}; on the open source side, we picked the mature OpenWhisk, due to its community support, constant evolution \cite{openserverless}, and backing of another popular commercial platform, \textit{IBM Cloud Functions}, with similar limits \cite{ibmcloud}. Not only are the numbers in Table~\ref{tab1:faas} far from the requirements above, the platforms' chaining best practices (OpenWhisk action sequences, \textit{AWS Step Functions}, and \textit{Azure Durable Functions}) are even more limiting, as intermediate dataframes can only be moved through object storage. Instead of the dataframe itself, functions return a placeholder, resulting in incorrect DAG semantics and sub-optimal performance (\S\ref{sec:passing}).

Given the limitations of existing FaaS platforms, are DAG tools coupled with traditional compute better equipped to satisfy these requirements? If we take the most popular framework, \textit{Airflow}, as an example, linking the programming model to \textit{any} kind of compute is a non-trivial operation for practitioners. Even when \textit{tasks} can be successfully coupled to a runtime (Kubernetes pods or an EC2 fleet), the framework does not provide dynamic scale-up, nor fine-grained, containerized management of Python dependencies. Importantly, \textit{tasks} are generic operators, not data-aware functions, leaving users to either implement their own data passing or rely on built-in, object storage backed primitives (``XComs''). \S\ref{sec:programming} below provides a vivid example of the shortcomings of this programming model when compared to \platx{}: unsurprisingly, the end result for the average data practitioner is that the learning curve is notoriously steep, and the developer experience sub-optimal \cite{Yasmin2024AnES}.

\begin{table}
\caption{Max. FaaS availability for I/O, memory, timeout}
\label{tab1:faas}
\begin{center}
\begin{tabular}{|l|c|c|c|}
\hline
\textbf{\textit{Platform}} & \textbf{\textit{Memory}} & \textbf{\textit{I/O payload}}& \textbf{\textit{Timeout}} \\
\hline
Lambda & 10GB & 6 MB & 900s  \\
Functions & 14GB & 100 MB & unlimited  \\
OpenWhisk & 2GB & 1 MB & 300s  \\
\hline
\end{tabular}
\end{center}
\end{table}

\section{Platform Design}

We now introduce \platx{}, a new FaaS service specialized for running data pipelines.  \platx{} is designed to scale up individual function invocations, efficiently pass large intermediate data, and allow users to interactively modify and run DAGs.  The rest of this sections is organized as follows: we enumerate the design principles behind \platx{} (\S\ref{sec:principles}), describe an architecture that prioritizes privacy of user data and deployment flexibility (\S\ref{sec:architecture}), and introduce a new programming model for developing data pipelines (\S\ref{sec:programming}).

\subsection{Design principles}
\label{sec:principles}

The goal of our FaaS system is to natively satisfy the three \textit{desiderata} in \S\ref{sec:whynew}: it is therefore convenient to state design principles which will allow us to operationalize those requirements. The biggest difference between \platx{} and existing tools is the conviction that achieving developer interactivity at the scale of modern data pipelines (\ie{}, hundreds of millions of rows moving across containers) is only possible if DAG abstractions and the runtime are co-designed: \ie{}, if serverless runtimes (\eg{}, \textit{AWS Lambda}) have no data awareness, and pipeline frameworks (\textit{Airflow}) have no runtime awareness, \platx{} provides both. Our \textit{user experience} principles are therefore a combination of architectural insights and data abstractions:

\begin{itemize}
    \item \textit{it runs in the cloud, but feels local}: as DAGs require moving GBs of data, only cloud bandwidth can guarantee a fast feedback loop. However, remote runtimes often come at the expense of developer convenience, with slow build times and no interactive logging (\S\ref{sec:container}).
    \item \textit{data awareness}: users should write code at the logical level of data dependencies, not at the physical level of data representation. In other words, user functions should specify as inputs \textit{only} tables, projections, and filters -- not files or endpoints. By restricting function signatures to (semantically specified) dataframes, we open the door for platform-level optimizations, such as aggressive caching, zero-copy data sharing, and versioning (\S\ref{sec:container}). 
\end{itemize}

Our \textit{system design} principles are focused on infrastructure and deployment:

\begin{itemize}
    \item \textit{ephemeral functions}: execution is truly stateless, as function instances live only for the duration of an invocation. In contrast, standard FaaS platforms typically reuse function instances across multiple invocations, making it harder for the average developer to reason about the program life cycle~\cite{10.1145/3360575}. Starting fresh each time simplifies reasoning about invocations, and it also allows instantly re-executing functions with different memory limits, which is a common requirement in data science (\eg{}, running a pipeline first on January data, then on the full year);
    \item \textit{off-the-shelf infrastructure}: cloud VMs provide the greatest level of customization and hardware diversity, which are both important levers when dealing with heterogeneous packages and large artifacts. VMs are also more portable than any serverless offering, appealing to enterprises who are sensitive to data security, and enabling \platx{} to support multiple deployment models in every major cloud, from a managed service (with or without a private link), to a full Bring-Your-Own-Cloud.
\end{itemize}

Taken together, our design principles are a considerable departure from FaaS standards: ephemeral existence, VM portability, and data awareness are necessary to satisfy data pipeline requirements, but these properties are less useful (and perhaps counterproductive) for traditional use cases involving relatively independent functions.

\subsection{Architecture}
\label{sec:architecture}

\begin{figure}
\centerline{\includegraphics[width=0.45\textwidth]{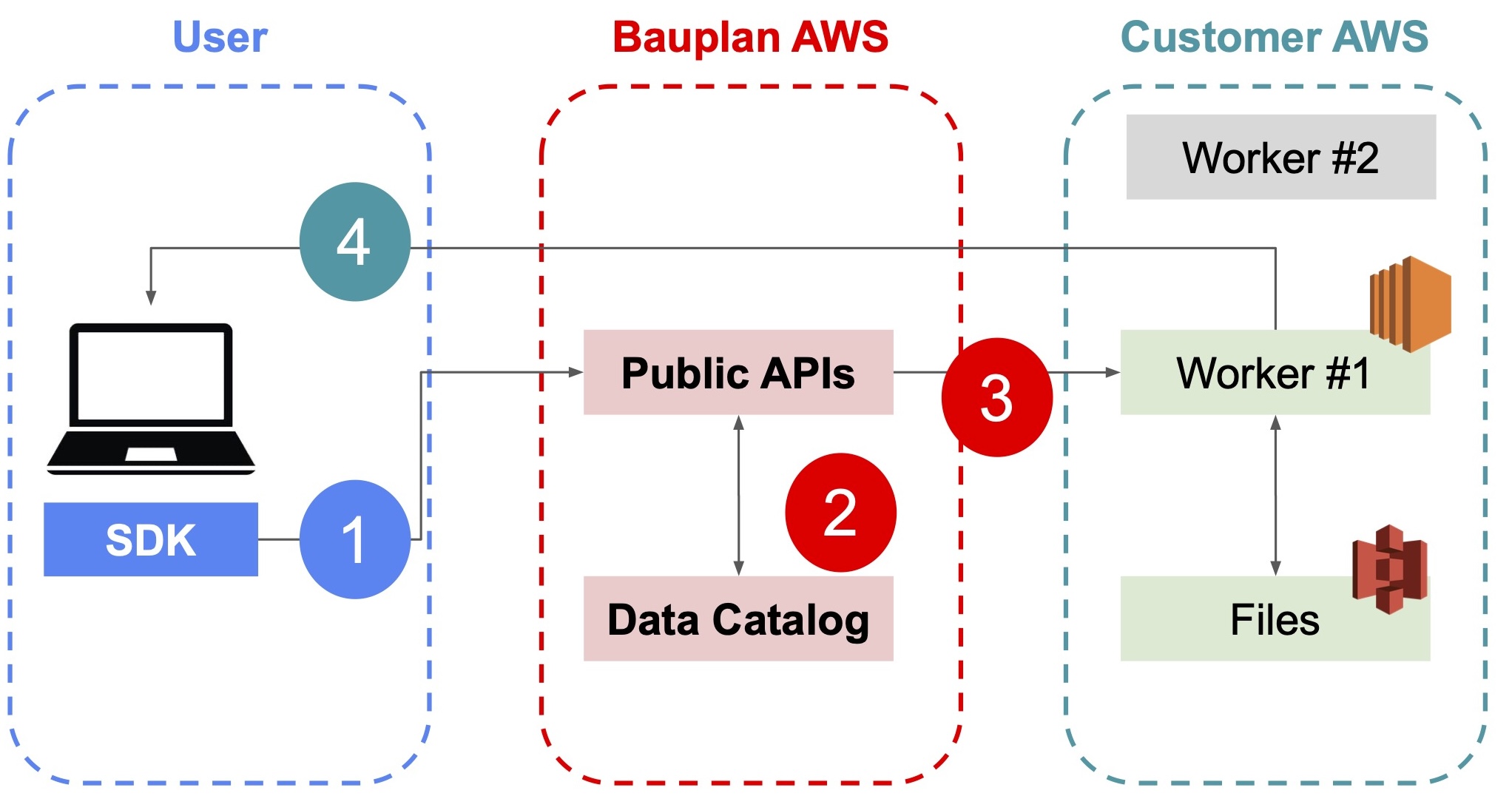}}
\caption{\textit{End-to-end architecture}: 1) A user requests to run a DAG; 2) the APIs parse the request and send an execution plan to a worker; 3) an existing (bin-packing) or on-demand worker runs the required operations over customer data inside the customer cloud; 4) print statements and data previews are streamed back to the user.}
\label{fig:bauplan}
\end{figure}

The basic modules involved in an \platx{} run are depicted in Fig.~\ref{fig:bauplan}. The system is built with a separation between the \textit{Control Plane} (CP) and the \textit{Data Plane} (DP):

\begin{itemize}
    \item the CP exposes multi-tenant APIs and only ever deals with \textit{metadata}: the CP is in \platx{}'s cloud account;
    \item for each customer we then have a single-tenant DP, a fleet of one or more off-the-shelf VMs which can be deployed (as in the picture) directly in a customer account. Workers are the only part of the system that interact with customer data; workers can be deployed in any cloud with ease as a Golang binary.
\end{itemize}

Crucially, the user and the workers are connected through bi-directional gRPC, so that every \texttt{print} statement in user code and system logs are visible in real-time in the user terminal / IDE. This is in stark contrast with platforms such as \textit{AWS Lambda}, which provide only asynchronous observability through Cloudwatch at additional cost. Debugging a run in \platx{} is a self-contained, real-time, and free operation; the execution feels as if it were local to the user.

\subsection{Programming model}
\label{sec:programming}

\textit{bauplan} provides a CLI tool and Python SDK installable with \textit{pip} (\ie{}, \texttt{pip install x}): the CLI allows users to initiate pipeline runs and navigate their data assets.  The SDK provides decorators for user code and a client to interact with the platform from any Python process. The best way to understand the developer experience is to see how the DAG in Fig.\ref{fig:dag} is expressed in \platx{}:

\addvspace{\baselineskip}
\begin{lstlisting}[showstringspaces=false,columns=fullflexible,language=Python,caption=A sample DAG in bauplan]
@bauplan.model()
@bauplan.python("3.11", pip={"pandas": "2.0"})
# the table name is the name of the function producing it
def euro_selection(
    # its parent node is referenced as the input
    data=bauplan.Model(
        "transactions",
        # columns and filters are expressed for 
        # pushdown to object storage
        columns=["id", "usd", "country"], 
        filter="eventTime BETWEEN 2023-01-01 AND 2023-02-01"
    ) 
):
   # do pre-processing here and return the 
   # cleaned dataframe directly
   return _df

# specify that the output needs to be written back to S3
@bauplan.model(materialize=True)
@bauplan.python("3.10", pip={"pandas": "1.5.3"})
def usd_by_country(
    data=bauplan.Model("euro_selection") 
):
   # aggregation code here - return as usual a dataframe
   return _df
\end{lstlisting}
\label{listing:bau}
\addvspace{\baselineskip}

Users express transformations as Python functions with the signature $f(dataframe(s)) \rightarrow dataframe$. The code is straightforward, but it is worth noting a few details:

\begin{itemize}
    \item the DAG topology is \textit{implicitly} expressed through function inputs;
    \item infrastructure is declarative: users specify the desired Python version and packages for each function, and the platform is in charge of deploying the corresponding stack -- two functions may use different interpreters and versions of \textit{pandas} within the same DAG;
    \item data management is declarative: users specify the desired input dataframe(s) for their code, and the platform is in charge of making it available inside the function by fetching it from object storage or from a parent. The optional hints on columns and filters enable optimizations such as predicate push-downs and columnar caching. Outputs follow the same principle: users return dataframes, and the platform is in charge of persisting them when required.
\end{itemize}

It is crucial to note that \platx{}'s declarative nature creates a principled division of labor between the system (infrastructure and data management) and the data scientist (business logic and choice of libraries). Furthermore, this decoupling is necessary to run the same pipeline over different versions of the same table (\eg{}, running today's code on last Friday's table \cite{10.1145/3650203.3663335}), or over different \textit{physical realizations} of the same asset (\eg{}, Parquet files in S3, Arrow streams over the wire, or locally cached data). In this regard, it is worth noting how \platx{} enables a true FaaS programming experience, in contrast to frameworks that instead couple the physical representation of a DAG with code.  For example, consider this \textit{AWS Airflow} reference implementation for a Python DAG \cite{awsairflow}:

\addvspace{\baselineskip}
\begin{lstlisting}[showstringspaces=false,columns=fullflexible,language=Python,caption=A pre-processing function in Airflow.]

def preprocess(
       s3_in_url, s3_out_bucket, s3_out_prefix
    ):
    # Do pre-processing here and save the result in
    # "s3_out_bucket / s3_out_prefix"
    return "SUCCESS"

# the function gets registered in the overall DAG
preprocess_task = PythonOperator(
    task_id="preprocessing",
    dag=dag,
    python_callable=preprocess.preprocess
)
\end{lstlisting}
\addvspace{\baselineskip}

Not only does the pre-processing function operate at the physical level of the raw data path instead of at a logical DAG level, but it saves its output as a side effect, instead of returning the cleaned dataframe to the caller. Hiding data artifacts from the scheduler prevents any further optimization, such as caching or compression.

\section{Implementation: the anatomy of a DAG run}

To understand how all of the pieces fit together, we will decompose a run and dive deeper in some of the system optimizations. To get a first-person perspective on the developer experience (real time log streams, function building, data passing, etc.), the reader is invited to pause here and watch a recorded run before continuing\footnote{Public video is available here: \url{https://www.loom.com/share/99ac0d5b5f944fc9aeef132bfaea0881}}.

A successful DAG run involves three main operations, which we will describe in more detail. First, translate user code to a physical plan.  User code is declarative, so the platform must fill the gap between \textit{logical} requests (\eg{}, ``I want \textit{transactions} with columns \textit{ID, USD, COUNTRY}'') and system operations (\eg{}, ``read files XYZ from S3'') (\S\ref{sec:logical+physical}).  Second, optimize environment construction so that the code can run in quickly provisioned ephemeral functions (\S\ref{sec:container}).  Third, optimize I/O and data movement (\S\ref{sec:passing}). In each step, \platx{} leverages data awareness and runtime awareness to perform optimizations not found in other comparable systems.

\subsection{Logical and physical plan}
\label{sec:logical+physical}

\begin{figure}
  \centering
  \includegraphics[width=\linewidth]{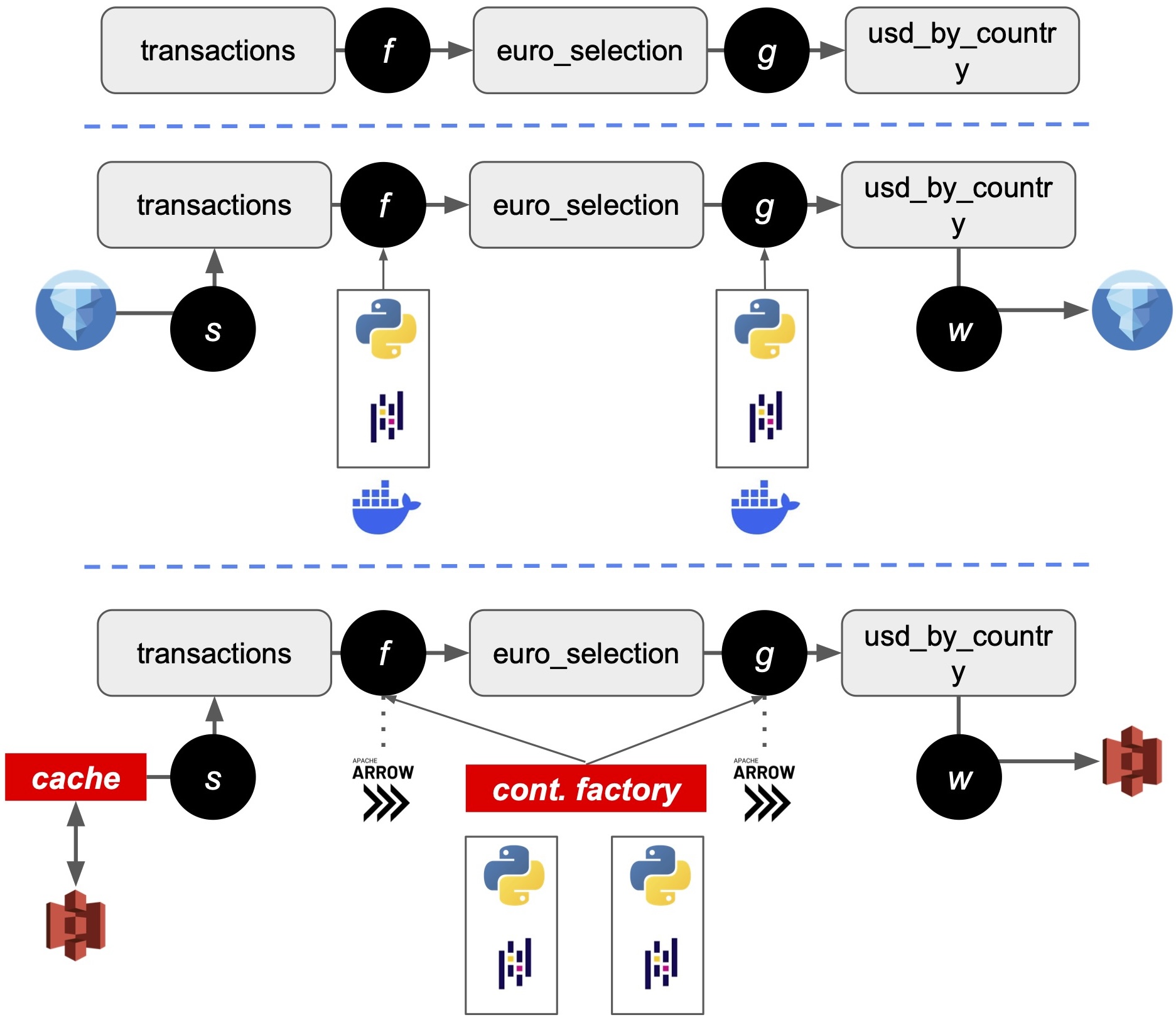}
  \vspace{-12pt}
  \caption{\textit{From logical plan to execution}, with three levels of representation for every run: (top to bottom) 1) logical plan, obtained by parsing user code, 2) physical plan, obtained by adding system functions, 3) the actual worker-level execution, transparently managing data and package caches to further speed up execution.}
  \label{fig:plan}
  \vspace{-10pt}
\end{figure}

Unlike other FaaS platforms, which execute user code ``as is'', \platx{} acts more like a database, where user code needs ``translation'' before being executed. When a run is requested from the CLI / SDK, the user code is uploaded to the control plane (CP) for processing and scheduling (Fig.~\ref{fig:bauplan}). The CP parses the code and reconstructs the DAG topology from the inputs / outputs of the functions: using database jargon, the result is the ``logical plan'', a structured representation of user code, expressing dependencies between steps with dataframe semantics (Fig.~\ref{fig:plan}, \textit{top}). This representation is too abstract to be sent to workers, so the CP will also produce a ``physical plan'' (Fig.~\ref{fig:plan}, \textit{middle}), which contains explicit instructions for the containerized runtime of the transformation functions \textit{f} and \textit{g} (\eg{}, \textit{euro\_selection} and \textit{usd\_by\_country}), and maps dataframe semantics to S3-backed tables. The CP accomplishes this by leveraging \textit{Iceberg} as its table format (providing schema evolution and per-table snapshots) and \textit{Nessie}~\cite{nessie} as a data catalog (providing cross-table transactions and data lake branching). Transparent to the user, the runtime has support for optimized procedures to read the \textit{transactions} dataframe from S3 and write back \textit{usd\_by\_country} as the final dataset. Throughout this procedure, no customer data is ever visible to the CP.

The physical plan is now specific enough to be sent to a worker (Fig.~\ref{fig:plan}, \textit{bottom}), which is in charge of executing it while taking into account local optimizations and constraints.\footnote{For simplicity of this exposition, we won't discuss the priority-based scheduler, which we plan to describe to the community in future work, and just assume the worker has full capacity to run the DAG at this point. Note that the programming model in Section~\ref{sec:programming} allows the same DAG to run on one or across multiple workers without code changes.}. In particular, the worker can access the files in the customer S3 bucket, as well as cleverly leverage space, time, and team locality: similar users with similar DAGs would access similar packages and S3 objects, opening up the possibility of performing significant optimizations in package (\S\ref{sec:container}) and data re-use (\S\ref{sec:passing}). Notably, all these optimizations require no code changes nor additional cognitive effort on the user side, and they would not be possible without \platx{}'s declarative programming model.

\subsection{Towards a fully interactive FaaS}
\label{sec:container}

Data practitioners frequently add or remove data science packages or update the versions of those packages. Like many other FaaS platforms, \textit{bauplan} guarantees code portability (``infrastructure-\textit{is}-code'') and selective maintenance: a new function can benefit from \textit{pandas==2.0} without forcing breaking changes in old functions with \textit{1.5.3}. However, due to its focus on interactivity and dynamic scaling, \textit{bauplan} significantly departs from existing FaaS along two dimensions: ephemeral function building and data caching.

As mentioned earlier (\S\ref{sec:principles}), functions only exist for the duration of a single invocation: two subsequent runs will build \textit{two} containers for the same \textit{euro\_selection} function (possibly with different hardware resources), which would result in unacceptable latency if re-building functions were a slow operation. To give a sense of the usual re-containerization flow for data science packages in popular platforms, Table \ref{tab:napkin} reports the latencies we recorded\footnote{Code snippets are available at \url{https://github.com/BauplanLabs/vldb-demo-2024}.} when re-running a DAG on a given target stack after adding a new Python package (the \textit{prophet} prediction library). \textit{bauplan} is 7x faster than \textit{Snowpark} -- a data-aware, serverless experience built within a cloud warehouse -- and 15x faster than \textit{AWS Lambda} at closing the developer loop, yet requires no additional tools, nor special knowledge. This improvement comes from two major insights. First, the worker leverages a local container factory to avoid re-installing common packages across runs, thus removing the dependency from network calls to PyPI. Second, our narrower use case compared to general platforms allows us to avoid relying on image layers (as for example \textit{AWS Lambda} does), with their associated building and network costs. In particular, our atomic building blocks for environments are the Python packages themselves.  To the best of our knowledge, we implemented OpenLambda-style package initialization~\cite{196322} in a Docker-compatible runtime as a novel experiment before any other research team. As a result, \platx{} containers are assembled in 100s of milliseconds by mounting relevant modules from the local file system.

Given all workers are single-tenant (Section~\ref{sec:architecture}), host disk and memory can be more easily shared between subsequent executions: instead of exposing APIs for manual management of temporary storage (like \textit{Lambda}), the \platx{} data cache works without user intervention, exploiting its signature with no side-effects. Once again, \platx{} data awareness makes database-like optimizations (not available in other platforms) possible:

\begin{itemize}
    \item intermediate dataframes are produced by DAGs functions, and \platx{} tracks both code and data changes, so it is possible to cache and re-use intermediate steps to avoid unnecessary re-computations when iterating;
    \item the semantics of reading \textit{Iceberg} tables follows relational algebra, so the cache can be \textit{columnar and differential}: for example, after reading from \textit{transactions} once, a subsequent request for \textit{ID, USD, COUNTRY, CLIENT\_ID} re-uses \textit{ID, USD, COUNTRY} from the cache and only downloads \textit{CLIENT\_ID} from object storage;
    \item inputs over object storage map to immutable files (through the \textit{Iceberg} manifest), so dataframe changes are identified with data commits such that the cache knows with certainty when a table is stale.
\end{itemize}

\begin{table}
  \caption{Time to add \textit{Prophet} to a serverless DAG}
  \label{tab:napkin}
  \begin{tabular}{lc}
    \toprule
    Task&Seconds\\
    \midrule
    \textbf{AWS Lambda}\footnote{As a best practice, Docker containers are run on \textit{AWS Lambda}, i.e. we re-build the container locally, ship it to ECR, launch CloudFormation to update the stack).} & \\   
    Update ECR container and function & 130 (80 + 50) \\
    \midrule
    \textbf{Snowpark} & \\ 
    Update Snowpark container& 35 \\
    \midrule
    \textbf{\platx{}} & \\ 
    Update runtime & 5 / 0 (cache) \\
  \bottomrule
\end{tabular}
\end{table}
 
\subsection{Intermediate dataframes}
\label{sec:passing}

As data workloads involve reading, writing, and moving around large dataframes, a \textit{cloud-only} design is a necessity, with cloud virtual machines (\eg{}, EC2 instances) offering network bandwidth up to 50 Gbps. However, running cloud functions with manual data management is still sub-optimal, as data scientists are likely to end up with slow implementations, low throughput, and complex logic to maintain. \platx{} constrains user inputs / outputs to be dataframes: raw dataframes in object storage are stored in Parquet files as \textit{Iceberg} tables, following lakehouse standards \cite{Zaharia2021LakehouseAN,Tagliabue2023BuildingAS} and maximizing interoperability. On the other hand, \platx{} represents intermediate dataframes as Arrow tables. Arrow is a popular open-source columnar format for in-memory and over-the-wire tabular data \cite{arrow}.  Arrow is built for modern CPUs and vectorized execution and has a fast-growing ecosystem built around it \cite{dremio, influx}. The Arrow format carefully avoids the use of pointers to memory addresses, preferring instead offset buffers to represent location and bitmaps to represent null values.  Avoiding pointers allows the same Arrow data to be mapped to different locations in different address spaces with minimal modification; as a result, Arrow avoids serialization and deserialization overheads and supports zero-copy transfers.  This results in significant performance gains relative to traditional columnar formats like Parquet or row-based wire protocols like JDBC. In the context of a DAG (such as the one in the demonstration above), Arrow is a crucial component for a fast feedback loop. 

As a pipeline is executed, the platform transparently picks a sharing mechanism: shared memory or local disk (for co-located functions) or Arrow Flight (across workers). To give a sense of the performance gains, Table~\ref{tab:reads} shows how long it takes to read an intermediate dataframe into a user function, depending on serialization and storage type\footnote{Code snippets are available at: \url{https://gist.github.com/jacopotagliabue/57bb14c675a5375338d4a57a88cea32a}.}; since existing platforms can only support S3-backed data passing (\S\ref{sec:whynew}), moving dataframes in \platx{} can \textit{be hundreds of time faster than alternatives}: perhaps counterintuitively, cross-machine communication through Arrow Flight is nearly as fast as local Parquet reading. Moreover, Arrow is very flexible: functions can transparently read tables from shared memory, memory-map them from disk (with standard OS primitives), or stream them from gRPC (with Flight) depending on resource availability.

Importantly, tables can often be shared with no movement at all: when children functions can be scheduled in the same worker, \platx{} performs a zero-copy sharing of the parent output. The Arrow IPC module allows children to transparently reference the underlying memory buffer, thus avoiding unnecessary copies: a 10 GBs table with three children only requires 10 (not 30) GBs, saving considerable time and resources.

\begin{table}
\caption{Reading a dataframe from a parent (\textit{c5.9xlarge}), avg. (SD) over 5 trials}
\begin{center}
\begin{tabular}{r|c|c}

 & \textbf{\textit{10M rows (6 GB)}}& \textbf{\textit{50M rows (30 GB)}} \\
\hline
Parquet file in S3& 1.26 (0.14) & 6.14 (0.98) \\
Parquet file on SSD& 0.92 (0.09) & 4.37 (0.15) \\
Arrow Flight & 0.96 (0.01) & 4.69 (0.01) \\
Arrow IPC & \textbf{0.01 (0.00)} & \textbf{0.03 (0.01)} \\

\end{tabular}
\label{tab:reads}
\end{center}
\end{table}

\section{Related Work}

Aside from platforms already discussed (\S\ref{sec:whynew}), there is academic interest in serverless systems that address some of the challenges we have encountered. For example, Mahgoub~\etal{}~\cite{273835} investigate different data passing mechanisms, but test only small inputs with no data-awareness; as Table~\ref{tab:reads} shows, tackling both storage \textit{and} formats is critical; Carreira~\etal{}~\cite{10.1145/3357223.3362711} provide orchestration for Machine Learning, using \textit{AWS Lambda} and \textit{Redis}, but focus only on training jobs; Jia~\etal{}~\cite{10.1145/3445814.3446701} also use shared memory to allow inter-function communication, but their stated goal is ``interactive microservices'', resulting in a low-latency, low-complexity system, with no affordances for data practitioners and no principled way to deal with storage and serialization trade-offs. Importantly, none of these systems is known to be used in production, making it hard to assess how the design choices actually fit real-world scenarios, especially given the known unreliability of analytics benchmarks~\cite{Renen2024}.

In general, data practitioners will be more familiar with DAG frameworks~\cite{airflow,prefect,luigi}; \textit{dbt}~\cite{dbt} pioneered the dataframe-based signature in the SQL community. While the DAG abstractions of these frameworks resemble \platx{}, they all leave data-awareness to the user: the result is often a tight coupling between business logic and physical data representation (\S\ref{sec:programming}), resulting in sub-optimal performance and  a developer experience far from the FaaS ideal.

\section{Conclusion}
While chaining is supported by major FaaS platforms, we have argued that their generality prevents them from fully supporting modern data pipelines. We outlined \platx{}, a new, purpose-built FaaS programming model and runtime, which effectively trades off control to achieve large gains in performance and developer experience. Notwithstanding its novelty, \platx{} is already used in production by large enterprises: as the platform further matures and more usage data are collected, we look forward to sharing with the community the next steps of our journey.

\bibliographystyle{ACM-Reference-Format}
\bibliography{sample-base}

\end{document}